# Theoretical-Experimental failure analysis of the c-Al$_{0.66}$Ti$_{0.33}$N-M2 steel system using nanoindentation instrumented and finite element analysis.

A.E. Gómez-Ovalle[1,4], M.Torres[2], J.M. Alvarado-Orozco[1,4], S.M.A. Jimenez[2], D.G. Espinosa-Arbelaez[1,4], J.M.Gonzalez-Carmona[1,4], J.Zárate-Medina[3], J. González-Hernández[1] and G.C. Mondragón-Rodríguez[1,4*]

[1] Center for Engineering and Industrial Development, CIDESI, Surface Engineering Department, Querétaro, Av. Pie de la Cuesta 702, 76125 Santiago de Querétaro, México.

[2] National Council for Science and Technology, (CONACYT), National Centre for Aeronautics Technologies (CENTA), Carretera Estatal 200, Parque Aeroespacial Querétaro, 76270, Colón, Querétaro.

[3] Metallurgy and Materials Research Institute, Universidad Michoacana de San Nicolás de Hidalgo, Av. Francisco J. Mujica, U Bldg., Ciudad Universitaria, 58030, Morelia Michoacán, México.

[4] Consorcio de Manufactura Aditiva, CONMAD, Av. Pie de la Cuesta 702, Desarrollo San Pablo, Querétaro, México.

*corresponding author: guillermo.mondragon@cidesi.edu.mx

**Abstract**

A theoretical-experimental methodology for failure analysis of the c-Al$_{0.66}$Ti$_{0.33}$N / Interface / M2 steel coating system is proposed here. This c-Al$_{0.66}$Ti$_{0.33}$N coating was deposited by the arc-PVD technique. For coating modeling the traction-separation law and the extended finite element method-XFEM were applied, the cohesive zones model was used for interface modeling and the Ramberg-Osgood law for substrate modeling. Experimental values using the instrumented nanoindentation technique, the scratch test and tensile stress test were obtained and introduced into the model. By means of nanoindentation the elastic modulus of coating, the fracture energy release rate and the nano-hardness. Normal and shear stress values of the interface were obtained with the scratch test, at the adhesive and cohesive critical loads. Vickers indentation was used to generate cracking patterns in the c-Al$_{0.66}$Ti$_{0.33}$N / Interface / M2 steel coating system. Radial and lateral cracks were generated and analyzed after transversal FIB cuts of the fracture zones. A finite element analysis was carried out to understand the relationship between the load-displacement curve and mechanical failure of in the system, associating the pop-in with nucleation, crack growth and cracking pattern. This works present a theoretical-experimental methodology for failure analysis of hard coatings (monolithic body) allowing to calculate fracture toughness of the coating material and model cracking patterns caused by contact mechanics.

Keywords: XFEM, cohesive zone model, fracture energy, c-Al$_{0.66}$Ti$_{0.33}$N-Interface-M2, lateral & radial cracks.

## 1. Introduction

Hard coatings have been widely applied to protect working tools in metalworking processes. Specifically, protective coatings are found on tools for machining operations of advanced alloys, *e.g.*, Ti- and Ni-base alloys, hard steels, and composite materials [1]. Additionally, process operations including high-speed dry machining, hot stamping, and extrusion are other examples of practical relevance [2]. In all these manufacturing processes, there is mechanical contact between a coated tool and working material. Under certain conditions, these operations are carried out between room temperature up to 1100 °C [2]. Upon mechanical contact, two failure modes have been identified between the coating-interface-substrate (CIS) system: i) the cohesive failures taking place in the bulk coating, and ii) adhesive failures occurring at the coating/substrate interface(s) [3]. For machining tools, the failure mechanisms of the CIS system is better described involving both damage responses, cohesive and adhesive fractures.

The answer to the question, how does each fracture mode contribute to the overall coating failure mechanism?, would shed light into the understanding of the mechanical response of CIS metal in working tool systems. However, experimentally there is no possibility to separate each contribution to the overall failure mechanism. Typically, experimental techniques based on contact mechanics, *e.g.*, instrumented nanoindentation and scratch testing are used to evaluate the mechanical response and failure mechanism of coatings systems [3]. Nanoindentation is the best state-of-the-art method to investigate the contact mechanical behavior of CIS systems. Based on the analysis of the indentation footprint and the load-displacement data, it is possible to calculate indentation hardness ($H_{IT}$), reduced elastic modulus ($E_r$) and fracture toughness ($\Gamma$) of the coating system [4]. Depending on the system evaluated, the load-displacement curve may show a discontinuity visualized like an extra-penetration at a constant mechanical load commonly known as "*pop-in*". This behavior has been associated with different phenomena including a pressure-induced phase transformation, the presence of microstructural defects such as voids, morphological changes, and stable crack growth prior to crack's blunt [5]. In the case of coating fracture, the *pop-in* events produced during the indentation load-displacement curve displays cohesive and adhesive combined effects. Moreover, in the *pop-in* regions is not possible to identify nucleation and fracture growth by conventional metallographic analysis, since cohesive and adhesive failures may interact making impossible to identify the nucleation site and the location where fracture start [6, 7]. An alternative approach to solve this contact mechanics problem involving failure mechanisms is combining experimental data and numerical methods, for instance, finite element method [8]. It is possible to build a model and numerically simulate the mechanical behavior of a CIS system in contact with an indenter taking advantage of experimental values and using them as inputs for the model [9]. This problem approach allows to model, independently or coupled, the behavior at the interface and the coating under certain contact mechanical loads.

An analysis of the interface between the coatings deposited on tools, employing the model of cohesive zones framed in the finite element method, has been already implemented by Xia et al. in [10]. The reported numerical model takes the energy of the interfacial fracture stress data as input parameters. To obtain these experimental values, a calculation of the area below the *pop-ins* zones from the indentation load- displacement curves has been applied. Applying this methodology, an interfacial fracture value of 2.18 MPa m$^{1/2}$ have been calculated by Ding et al. for TiN-interface-steel systems [11]. These experimental values can be used as input parameters in the traction-separation function, to model the energy or critical stress required to cause an interfacial separation, e.g., a coating delamination [12]. The traction-separation law allows modeling the interfacial energy release, crack nucleation, and interfacial crack growth related to each failure mechanism.

More recently, the extended finite element method (X-FEM) has been used to model the cracking behavior of CIS systems [13]. X-FEM method allows modeling crack growth, without the restriction to re-mesh the system geometry, by introducing discontinuous enrichment functions increasing the nodes degrees of freedom [14]. X-FEM method also allows to implement criteria functions obtained from analytical solutions to predict crack growth directions [15]. In this way, it is possible to model more freely the crack growth within a continuous region of the CIS system. For instance, Csanádi et al. [16] modeled the crack propagation in a WC-coating deposited on tool-grade steels using X-FEM reporting the relationship between the cracking pattern and sink-in and pile-up phenomena [17]. These results were in good agreement with experimental nanoindentation data [18]. Therefore, the combination of the cohesive zones and the X-FEM methods have proven to be useful for better understanding of failure mechanisms on hard coatings. In other words, these methods allow to investigate and separate failure contributions and therefore identifying probably nucleation sites and crack growth [19]. Additionally, cracks can nucleate and grow inside the bulk of the coating, making its *in-situ* identification challenging and as an alternative to investigate these phenomena is through the use of numerical models and its comparison with *post-mortem* analysis of the coating system.

In this research work, a theoretical-experimental approach to investigate the failure analysis of a c-$Al_{0.66}Ti_{0.33}N$ / Interface / M2 steel coating system is proposed. Failure behavior of the CIS system under normal mechanical contact at ambient temperature is analyzed and discussed. The mechanical behavior, crack growth and failure pattern described by the numerical modeling calculations applying the cohesive zone model and X-FEM method are compared with experimental nanoindentation results.

## 2. Methodology

The theoretical-experimental methodology proposed in this work contains three steps: (1) experimental set up: first c-$Al_{0.66}Ti_{0.33}N$ was deposited on M2 steel substrates applying arc-PVD technique, followed by micro-structural and nano-mechanical characterization; then (2) proposal of the finite element model based on the experimental data, and (3) analysis of the failure mechanisms of the c-$Al_{0.66}Ti_{0.33}N$/interface/M2 system.

### 2.1 Experimental setup

**Coating deposition process**

The c-$Al_{0.66}Ti_{0.33}N$ coatings were deposited on mirror polished AISI M2 steel substrates (Ra = 100 nm) using a PVD Coater unit model Domino Mini from Oerlikon Balzers. The cathodic arc PVD technology using a 99.5 % pure Ti/Al (33/66) targets from Plansee was used. During the deposition process, a base working pressure of 3 mTorr, a cathode amperage hour of 400 Ah and a graded bias voltage from -40 to -80 V were defined. The final thickness obtained for the coating was ~ 3.8 µm.

**Mechanical properties of M2 steel substrate**

The mechanical properties of M2 steel substrate were measured by means of the tension test under the E8/E8M - 16a standard [20]. The straining rate was 0.015/6/0.006 mm/mm/min [in./in./min]. For this purpose, cylindrical rod specimens were machined according to the standard pattern number 1, and after testing, the elastic modulus, strain hardening index and yield resistance were calculated.

**Crystalline structure and crack pattern of the coating**

Crystalline structure of the arc-PVD c-$Ti_{0.33}Al_{0.66}N$ coating was determined by X-ray diffraction using a Rigaku Smartlab equipment. The grazing incidence technique was applied to allow diffraction of the thin-film coating. An X-ray source of Cu kα with a wavelength λ of 1.54 Å and nickel filter were used. X-ray data were obtained at 0.2° incidence angle, in a 2θ range between 10° and 80° with a step size of 0.020°. The phase indexation and the lattice parameters of the crystal structure were obtained through the PDXL software and the ICDD database. The microstructure and crack pattern of the coatings was analyzed by cross-sectional transmission electron microscopy (S-TEM) using 200 KeV energy in a JEOL JEM 2200FS+CS.

**Mechanical evaluation of the coating-substrate system**

The mechanical properties were measured by means of the instrumented nanoindentation technique thus obtaining $H_{IT}$, $E_r$ and Γ along with fracture characteristics of the c-$Ti_{0.33}Al_{0.66}N$ coating. The nanoindentation tests were carried out in a Nanovea indenter for hardness profile at high loads (5 mN - 300 mN) and a Hysitron for small loads (0.6 mN - 5 mN). In both cases, a Berkovich diamond indenter was used, applying constant loading-unloading cycles under load control mode. The load *vs* displacement (*P vs h*) curves were recorded within 20 seconds and analyzed by the Oliver and Pharr method [21]. Corrections due to the effects of thermal drift, creep and equipment compliance were considered in the test. Prior coating analysis fused silica was measured as a reference. The Poisson's ratio (ν), $E_r$ and $H_{IT}$ of this sample are summarized in Table I. The experimental values suit well the information provided by the Nanoindentation supplier. Nanoindentation test on the c-$Ti_{0.33}Al_{0.66}N$ coating was designed to obtain the $H_{IT}$ profiles and $E_r$ at the following loads: 1 to 5, 10, 15, 20, 30, 50, 70, 100, 150, 200, 250 and 300 mN. The statistical analysis of the data was carried using the Weibull method [22]. Vickers indentations were done on the coatings surfaces to induce cracks using a 5 N load. The indentation footprint were analyzed with a Jeol Neoscope jcm-500 scanning electron microscope, both superficial and transverse by means of Focused Ion Beam (FIB). Finally, a roughing polishing of a 20 x 20 um section was performed to reveal the microstructure of the film in the JEM9320-FIB system using a Ga + ion beam.

**Table I. Mechanical properties of mirror polished Fused silica.**

| Reference | E (GPa) | $\nu$ | H (GPa) |
|---|---|---|---|
| Fused silica | 72 ± 0.089 | 0.17 | 8.989 ± 1 |

Dynamic scratch testing was carried out to evaluate coating adhesion. The critical load of the CIS system, which is considered a measure of the interfacial resistance, was determined from the drag coefficient (CoFA) *vs* distance graph as suggested in the ASTM 1624C-05 standard [23]. The scratch test conditions were: Rockwell C indenter, 0.1 to 35 N load gradient, scratch speed of 14 N/min, 3.5 mm stroke and travel speed of 11 mm/min. Failure modes and scratch tracks footprints were characterized and compared with the ASTM C1624C-05 standard [23]. Data was obtained and analyzed using the Nanovea Scratch software, and the scratch tracks were characterized using a scanning electron microscopy JEOL JEM 2200FS+CS.

**2.2 Modeling and finite element simulations.**

Fig. 1 shows a schematic of the CIS system and the constitutive models used to represent its mechanical/failure behavior. The normal contact between the cutting tool and the chip is approximated by an instrumented nanoindentation test. In this model, the chip is represented by a symmetrical spherical indenter. For the numerical model, the spherical indenter is considered effectively rigid in comparison to the c-$Al_{0.66}Ti_{0.33}N$ coating and the M2 steel substrate. Extended finite elements are used to discretize the coating. The asymptotic crack-tip functions associated with the enriched/singular degrees of freedom are given by the Westergaard function. On the other hand, M2 steel substrate is assumed to be an elastoplastic material governed by the Ramberg-Osgood constitutive law. In this manner, localized crack nucleation and propagation along the interface is permitted without the introduction of pre-existing cracks into the model. Note that the cohesive zone response is defined by specifying the normal and tangential tractions ($T_n$, $T_t$) acting on their conjugate separations ($d_n$, $d_t$), respectively.

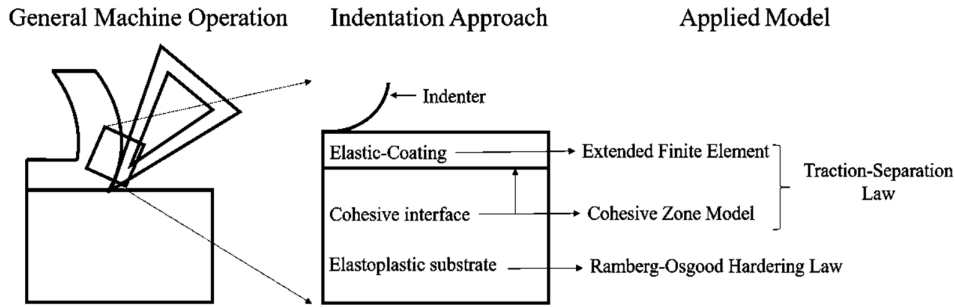

Figure 1. General schema of the coating / interface / substrate system and modelling strategy.

The commercial software ABAQUS® was used to build and simulated the numerical models described in this section. 3D Finite Element Method considering an elastic spherical indenter in contact with c-$Al_{0.66}Ti_{0.33}N$ / Interface / M2 steel coating system with thicknesses of 3.8/0.001/200 μm respectively was built to model the nanoindentation test.

**Description of the applied models**

The extended finite element method uses the following approximation function to model crack growth [24]:

$$u^h(x) = \sum_{I \in N_I} N_I(x)u^I + \sum_{J \in N_J} N_J(x)H(x)\,a^J + \sum_{K \in N_K} N_K(x) \sum_{\alpha=1}^{4} f_\alpha(x)\,b^{K\alpha} \quad (1)$$

where $u^h$ corresponds to the displacement vector, the functions $N$ correspond to the shape functions, $u^I$ corresponds to the nodal displacement vectors, $H$ corresponds to the jump Heaviside function, $f_\alpha$ corresponds to the asymptotic crack-tip functions and $b$ corresponds to nodal enriched degree of freedom vector. The first term on the right side is related to the standard shape function of the conventional finite element method. The second term is related to the discontinuous enrichment generated by the Heaviside function. The third term is related to the singular enrichment at the tip due to analytical solutions that predict the direction of growth of the crack.

Fig. 2 shows the separation traction law which models the growth of new surfaces due to the cracks nucleation and their growth. This law is applied in the cohesive zones model and in the extended finite element method. In this case, there are three important parts: the first corresponds to the damage initiation point ($\sigma_{max}$, $\Delta_c$) at the maximum stress and critical displacement, the second part corresponds to the line with a negative slope that corresponds to damage evolution and the third part corresponding to the final stage of failure.

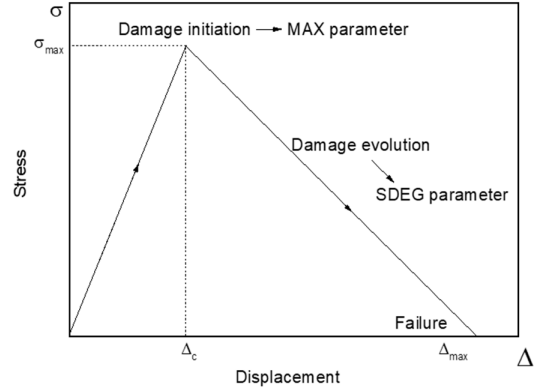

Figure 2. Bilinear function that represents the criteria of failure used in the traction separation law used in the model of cohesive zones and in the extended finite element method.

The criteria for the interface damage initiation is defined by the MAXS parameter [24], that is, assuming that the damage is initiated when the ratio of maximum nominal stress approaches to 1, according to equation 2:

$$max\{\frac{\langle t_n \rangle}{t_n^0}, \frac{t_s}{t_s^0}, \frac{t_t}{t_t^0}\} = 1 \quad (2)$$

where $t_n^0, t_s^0$ and $t_t^0$ represent the peak values of the nominal stress when the deformation is purely normal to the interface or purely shear in the first or second direction, respectively. The symbol '<>' used in the equation 2 represents the parentheses of Macaulay used to describe that a pure compressive deformation or its stress counterpart does not initiate damage.

The damage evolution criterion is based on an energy term [24], with the form of a power law:

$$\{\frac{G_n}{G_n^C}\}^\propto + \{\frac{G_s}{G_s^C}\}^\propto + \{\frac{G_t}{G_t^C}\}^\propto = 1 \quad (3)$$

The mixed mode of fracture energy is defined as $G^C = G_n + G_s + G_t$ when the previous condition is satisfied. In equation 3, the quantities $G_n, G_s$ and $G_t$ refer to the work done by the traction and this conjugate relative to the displacement in the normal direction, and in the first and second shear respectively. The amounts refer to the critical fracture energies required to cause failure in these three directions respectively. For simplicity, it was assumed that the power equals $\propto = 1.0$ [24]. Therefore, the damage evolution criterion becomes:

$$\frac{G_n}{G_n^C} + \frac{G_s}{G_s^C} + \frac{G_t}{G_t^C} = 1 \quad (4)$$

For the FEA simulations, C3D8R (8-node linear brick, reduced integration with hourglass control) elements were used. Three finite element models were constructed with 40, 50 and 60 elements at the contact radius (half of the contact). The element sizes were 0.40585, 0.31341 and 0.24398 μm respectively. The convergence analysis was carried out at 20, 30, 40, 50 and 60 mesh elements in contact. The calculated contact area between the indenter and the system surface was determined for each mesh and its variations are observed in the Fig. 3.

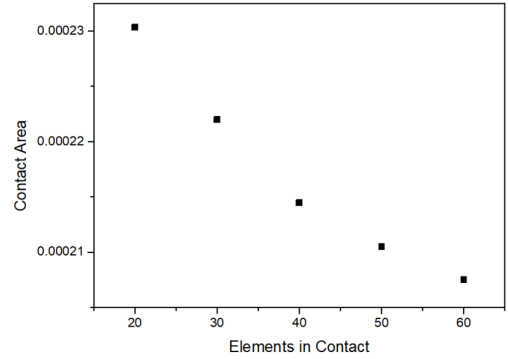

Figure 3. Convergence analysis by means of the contact area of the Indenter-Surface.

**2.3 Development of the finite element model based on the experimental data**

Fig. 4 shows the relationship between the mechanical characterization and the analytical and numerical models developed and implemented in this work.

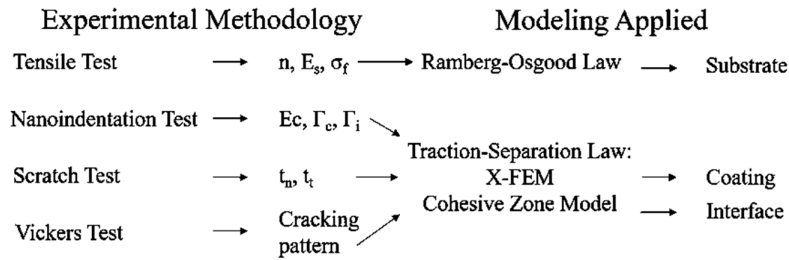

Figure 4. Schema of the theoretical-experimental methodology applied in this work.

Thus, the hardening coefficient (n), elastic modulus ($E_s$) and yield strength ($\sigma_f$) of the substrate were obtained from the tensile test. $E_s$ corresponds to the slope obtained by fitting with a linear function the elastic part of the stress-strain curve. $\sigma_f$ is calculated considering the cut of the line parallel to the elastic part of the curve with the curve stress-strain, considering a deformation corresponding to 0.02% . Regarding the elastoplastic part of the curve, the fitting with power law gives the hardening coefficient corresponding to n. These properties are input parameters for the Ramberg-Osgood Law, which has been applied to model the elasto-plastic behavior of the substrate.

The instrumented nanoindentation technique was used to calculate the $E_r$, fracture energy ($\Gamma_c$) and interfacial fracture energy ($\Gamma_i$) of the c-$Ti_{0.33}Al_{0.66}N$ coating. $\Gamma_c$ is calculated by means of the values obtained by nanoindentation and the model of the indentation-driven channel cracking [25] considering our experimental parameters, the radial cracks generated by the Vickers test (eq. 5).

$$K_I = \Gamma c = \lambda V^* \frac{(E_c H_c^2)^{1/3}}{t_c} \frac{a^2}{c^{1/2}} \qquad (5)$$

In this case, $\lambda$ is approximately 0.013 for the Vickers geometry, $a$ is the diagonal of the indentation, $c$ is the length of the crack. $E_c$ and $H_c$ correspond to the elastic modulus and hardness of the coating respectively. The parameter $V^*$ is related to the plastic deformation generated in the substrate due to indentation,

$$V^* = \begin{cases} 1 & h_d \leq t_c \\ 1 - \frac{(h_d - t_c)^3}{h_p^3} & h_d > t_c \end{cases} \qquad (6)$$

where $h_d$ is the indentation depth. On the other hand, $V^*$ is related to the elastic modulus to determine $\Gamma_c$ according to equation 5. For its part, $\Gamma_i$ was calculated by considering an energy balance between the work done by the indenter ($\delta W$), the elastic deformation energy ($\delta U_e$) of the coating-substrate system, the associated plastic dissipation during crack propagation ($\delta W_p$), the energy for interfacial ($\delta U_i$) and cohesive ($\delta U_c$) crack extension [26]. In this way, the conservation of energy determines that:

$$\delta W = \delta U_e + \delta W_p + \delta U_i + \delta U_c \qquad (7)$$

The energy terms corresponding to $\delta U_i$ and $\delta U_c$ are related to toughness in the following way:

$$\delta U_c = \Gamma_c \delta a_c \qquad (8)$$

where $\Gamma_i$ y $\Gamma_c$ can also be defined as the fracture toughness of adhesive and cohesive cracks respectively. On the other hand, $a_i$ and $a_c$ correspond to the crack increase zone in the *pop-in*. In this way, $\delta W$ is measured directly from the load-displacement curve, calculating the area under the curve in the *pop-in* region. The $\delta W$ value involves energies which are related with the crack extension and only associated to the normal loading applied during indentation. Therefore, this is an approximate value for the energy needed for crack extension in the system. The scratch test was used to measure the following parameters: critical failure stresses, $T_n$ (normal stress) and $T_t$ (tangential stress). These values are obtained from the critical loads (adhesive and cohesive) and considering the approximate area of the indenter, which has a radius of 200 µm and spherical geometry.

## 3. Results and Discussion

### 3.1 Experimental data

**M2-steel substrate**

To obtain the mechanical properties of the substrate, the tensile test was carried out. Fig. 5 shows the stress-strain curve of the M2-steel substrate. In this case, two types of fitting to the curve were made, a linear fit to the elastic zone and a fit with power law for the elastoplastic zone. Thus, Fig. 5 shows the fitting equations for these two regions. For the elastic part, the slope of the linear function corresponds to the value of the $E_s$ = 185.7 GPa. Yield stress was calculated considering a line parallel to the elastic part of the curve with a deformation corresponding to 0.02 %, in this case the $\sigma_f$ = 386.6 GPa. Regarding the elastoplastic zone of the curve, fitting with power law generates the coefficient corresponding to the hardening coefficient, this being 0.23. These results are used as inputs parameters for modeling the elastoplastic behavior of the substrate.

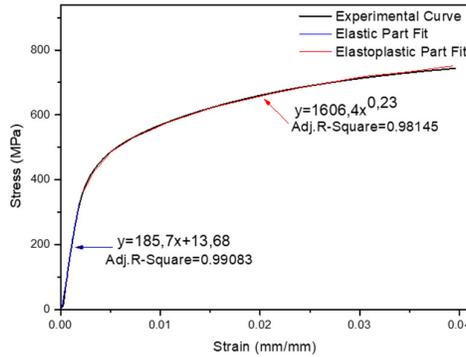

Figure 5. Tensile test curve of AISI M2 steel.

**Coating characterization**

Fig. 6 shows the grazing incidence X-ray diffraction pattern of the as-deposited coating confirming the cubic structure of the solid solution $Ti_{0.33}Al_{0.66}N$ according to JCPDS Card: 01-071-4029. The coating showed a preferential orientation in the (200) direction and no other crystallographic phases such as TiN or AlN were observed.

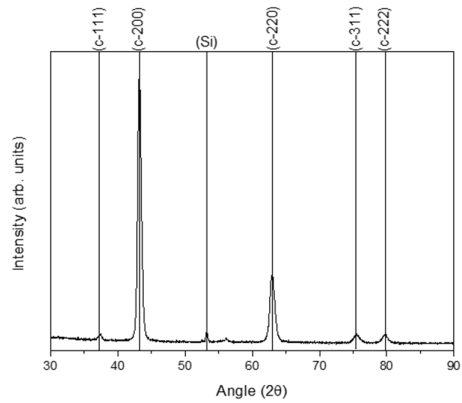

Figure 6. Grazing incidence XRD pattern of the as-deposited arc-PVD $Ti_{0.33}Al_{0.66}N$ coating.

In Fig.7, a STEM image of the cross section FIB cut of the arc PVD c-$Ti_{0.33}Al_{0.66}N$ coating is showed. Typical defects such as droplets, associated to the deposition technique arc-PVD, can be identified. Diffusion zones can be elucidated at the coating and the substrate interface.

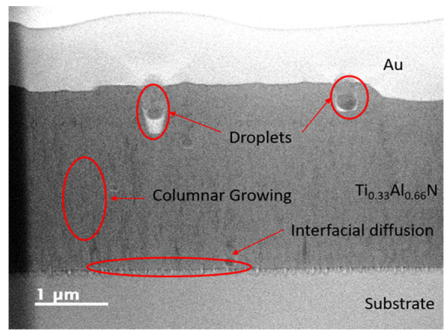

Figure 7. STEM image of a cross-section of the as-deposited arc-PVD c-$Ti_{0.33}Al_{0.66}N$ coating.

**Mechanical characterization of the CIS system.**

*Nanoindentation experiments*

The elastic modulus ($E_r$) and indentation hardness ($H_{IT}$) of the coating, measured by nanoindentation were 402 GPa and 32 GPa respectively. An indentation hardness profile as a function of indenter relative displacement ($h/t$) was made and the corresponding indentation prints were analyzed by SEM. The composite hardness profile made on as-deposited arc-PVD c-Ti$_{0.33}$Al$_{0.66}$N / interface / M2 steel coating system is shown in Fig.8. In the plot, three characteristic zones can be seen: the first region corresponds to the coating hardness (*i.e.*, displacements about 10 % of the coating thickness), the second zone corresponds to hardness of coating + substrate, and the third region corresponds exclusively to the substrate response. Steel hardness was 4.6 GPa.

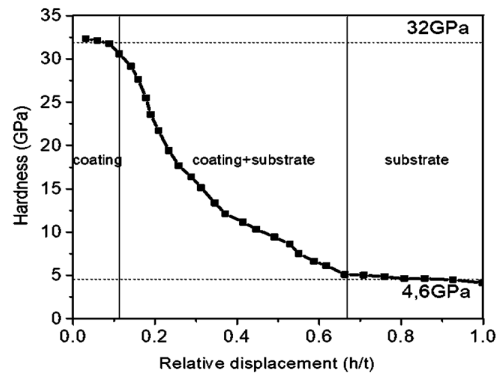

Figure 8. Evolution of the indentation hardness ($H_{IT}$) as a function of indenter relative displacement ($h/t$). Indentation hardness profile values of the CIS system.

Fig. 9-left shows the load-displacement curve of an as-deposited c-Ti$_{0.33}$Al$_{0.66}$N coating system at maximum 300 mN load, showing an onset failure load of the coating at about 150 mN. Micrographs of the indentation prints at 200 and 300 mN indicated that the c-Ti$_{0.33}$Al$_{0.66}$N coating system failure mechanism is mainly characterized by plastic deformation. On the other hand, no radial, lateral or medium cracks (i.e., cohesive and adhesive failures) were observed on the surface of the coating at these loads levels (see Fig 9-right). However, two *pop-in* zones were clearly identified in Fig. 9-left which are associated with irreversible processes such as coating fracture and delamination [27]. The indentation work $\delta W_1$ y $\delta W_2$ done by the indenter to generate the crack are observed in Fig. 9-left. These are calculated by means of the area under the curve generated by the *pop-in*.

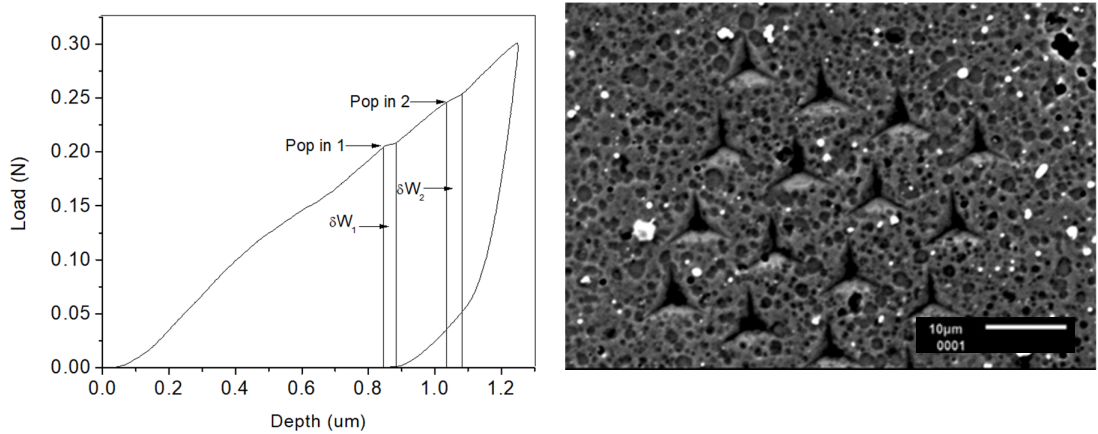

Figure 9. (left) Load *vs* displacement curve of c-Ti$_{0.33}$Al$_{0.66}$N coating at 300 mN. Two *pop-ins* and work regions of the indenter to start cracking are observed, and (right) nanoindentation prints after 300 mN load on the c-Ti$_{0.33}$Al$_{0.66}$N coating surface.

*Vicker´s indentations for coating failure and crack evaluation*

In order to promote the coating failure and induce cracking, Vickers indentations applying 5 N load were carried out on the surface of the c-Ti$_{0.33}$Al$_{0.66}$N / Interface / M2 system. Radial and lateral cracks (L-cracks) were observed at the indentation prints (Fig. 10a). The presence of radial cracks (R-cracks) are related to cohesive failures due to stress concentration in the corners of the indenter, and these cracks are produced during indentation unload. On the other hand, lateral cracks nucleate and grow due to the high-stress concentration at the edges of the indentation as the indenter penetrates appearing during indentation loading [10]. To analyze the fracture characteristics of the c-Ti$_{0.33}$Al$_{0.66}$N / Interface / M2 system on these two failure regions, FIB cuts were taken as labeled in Fig. 10a. CS1 cut corresponds to lateral crack (Fig. 10c) and CS2 cut is related to the radial crack Fig. 10b. Considering lateral cracking (L-crack), three types were identified. The first L-crack type (a-insert in Fig. 10c) is a lateral crack, which nucleates on the coating top surface and grows perpendicular to the surface. Crack opening at the coating surface is also observed. The second L-crack type (b-insert in Fig. 10c) appears as possible nucleation site where tortuous cracking occurs crossing the interface and reaching the substrate. The third internal L-crack type nucleates at the interface and grows inside the coating but does not reach the open surface. Radial crack (R-crack) nucleates on the coating surface and grows perpendicular into it. In a similar way, it is observed that the R-crack crosses the interface and when it reaches the substrate the R-crack is deviated. Crack opening is not as pronounced as in the lateral crack observed in Fig.10c.

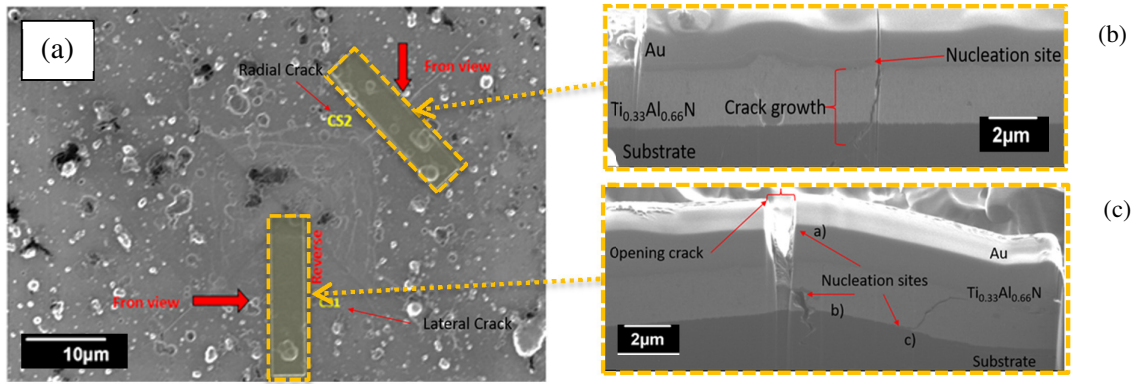

Figure 10. a) 5 N Vickers indentation footprint, b) lateral crack FIB cut, c) radial crack FIB cut.

*Scratch testing*

Fig. 11 shows the friction coefficient *vs* scratch distance of the as-deposited c-Ti$_{0.33}$Al$_{0.66}$N coating. Three critical loads were determined (Lc1, Lc2 and Lc3). Lc1 corresponds to the critical failure cohesive and Lc1 together with Lc2 are related to adhesives failures. The critical adhesive loads are generated in the last sections of the distance, which shows the good adhesion of the coating. The above corroborates what was observed in the nanoindentation fingerprints. The good adhesion of the coating is justified with the diffusion zone observed in the microstructures. This diffusion zone decreases the interfacial discontinuity between the coating and the substrate. Coating failures induced by scratch tests were analyzed by SEM (Fig. 12a). Lateral cracks (Fig. 12b) are observed in the scratch track which correspond to cohesive failures associated with LC1. Chevron cracks (Fig. 12c) are observed at the end of the scratch track, which are related to adhesive failures due to the general delamination of the coating, and therefore correspond to LC2. The LC3 critical load is related to the material accumulation in the very last sections of the scratch track due to the plastic deformation of the substrate. The adhesive critical load LC2 was used to calculate the interfacial failure stresses that were implemented in the failure models.

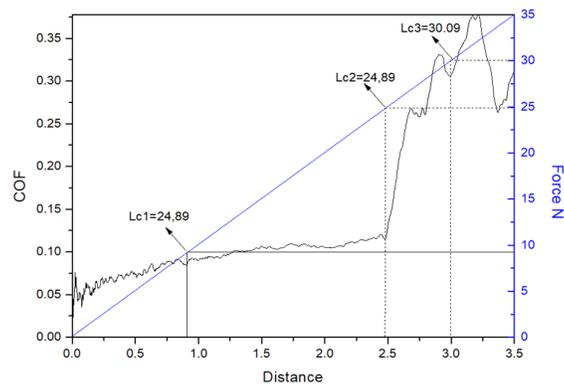

Figure 11. COF vs distance curve generated by scratch test. Three critical loads are observed.

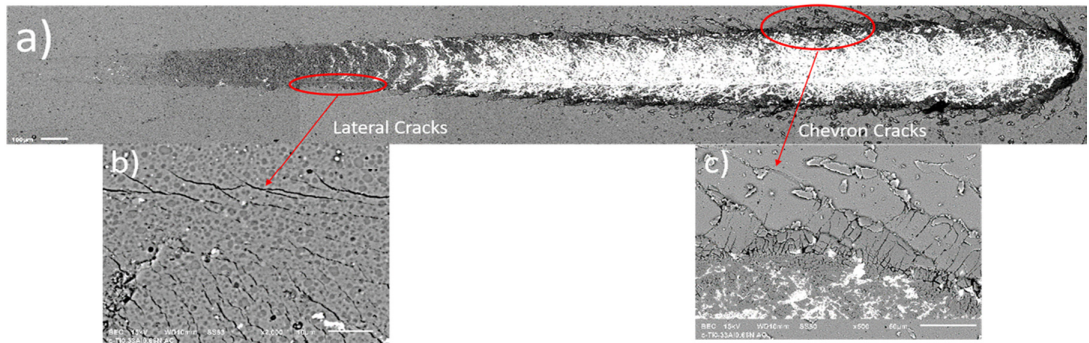

Figure 12. Scratch testing of the c-Ti$_{0.33}$Al$_{0.66}$N arc-PVD coating. a) Complete wear track. b) Chevron cracks on the edges of the wear track (. c) Lateral cracks on the edges of the wear track

Table II summarizes the experimental mechanical Properties of the Coating / Interface / Substrate System. This table contains the experimental method applied for each system element and their property values used to XFEM model the CIS system and to analyze their theoretical-experimental failure mechanisms as shown in the next section.

**Table II. Mechanical properties of the Coating / Interface / Substrate system. Data applied in the FE Models.**

| System element | Property | Value | Experimental Method |
|---|---|---|---|
| AISI M2 steel Substrate | n | 0.23 | Tensile Test |
| | $E_s$ | 185.7 GPa | |
| | $\sigma_y$ | 0.3866 GPa | |
| | $\nu_s$ | 0.29 | Data from [28] |
| c-Ti$_{0.33}$Al$_{0.66}$ arc PVD coating | $E_c$ | 402 GPa | Nanoindentation |
| | $\Gamma_c$ | 120 J/m2 | |
| | $\nu_c$ | 0.2 | Data from [29] |
| Interface | $\Gamma_i$ | 83 J/m2 | Nanoindentation |
| | $t_n$ | 2.088 GPa | Scratch Test |
| | $t_t$ | 0.299 GPa | |

## 3.2 Numerical model of the Coating / Interface / Substrate (CIS) nanoindentation.

The cohesive zones model was implemented to simulate the interface as an independent system-element. For this purpose, the critical load value from the scratch test was used as a parameter related to the stress fracture. Fig. 13 shows the delamination of the CIS system. This result is important, since it shows that the calculation obtained for the interfacial fracture energy$\Gamma_i$, is an adequate value to predict the delamination in this type of systems. This is corroborated by the fact that experimentally (figure 10c) interfacial cracks were obtained in a similar state of load.

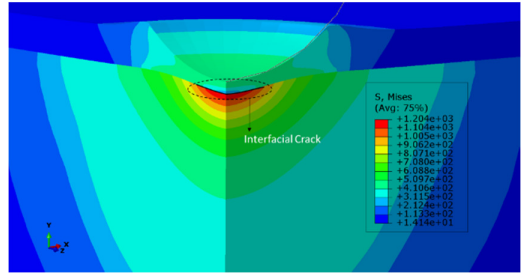

Figure 13. Stress field considering the interface in the FE-modelled system.

Fig. 14 shows the failure evolution during the first stages of the indentation. In Fig. 14a, the beginning of the contact between the indenter and the coating is observed. The nucleation of a lateral crack at the edge of the contact between the indenter and the coating is observed in Fig. 14b. In the Fig. 14c the growth along the thickness of the coating with cone-like geometry is observed. In the Fig. 14d the opening of the crack is observed.

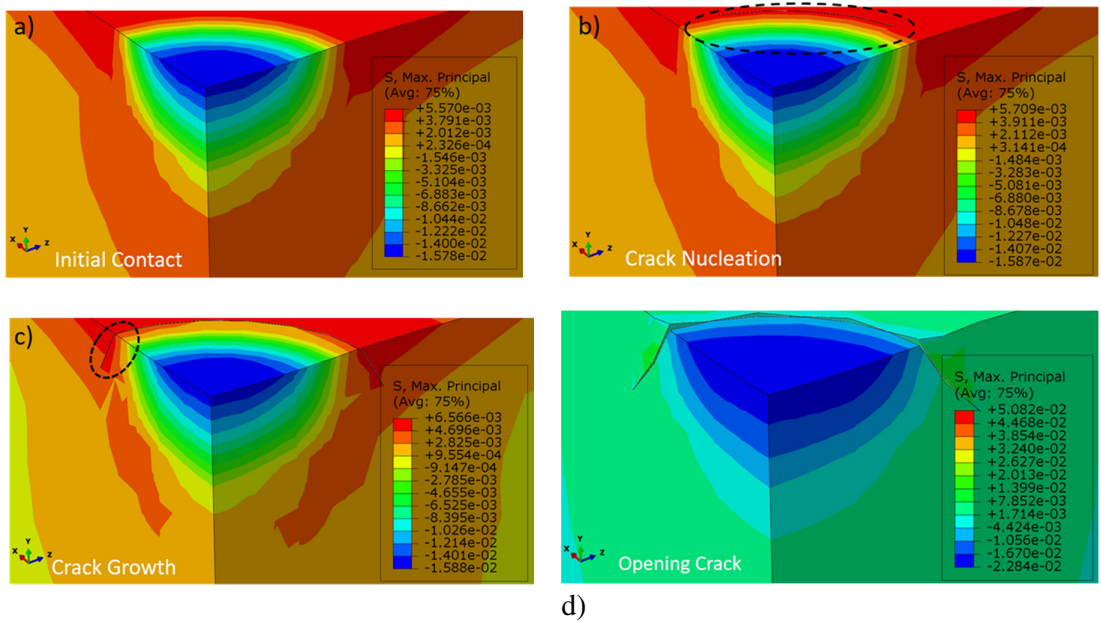

d)

Figure.14 Evolution of failure in the analyzing stress maximum principal.

Fig. 15 shows two load vs. displacement curves, the first curve is obtained applying the cohesive zone model and the second is based on a model including the cohesive zones and the extended finite element method (XFEM). The load vs. displacement curve without XFEM has a larger area under the curve. This is because the work exerted by the indenter in the model of cohesive zones is transformed into the energy necessary to generate an interfacial crack plus plastic deformation. On the other hand, when the model of cohesive zones is applied together with X-FEM, this work, in addition to being used to generate the interfacial crack, is used to generate the cohesive crack. Therefore, the effect due to plastic deformation is relatively lower. The P-h curves allow to determine that cohesive crack nucleates before the interfacial crack. It is important to note that experimentally the same crack types with similar growth pattern were obtained.

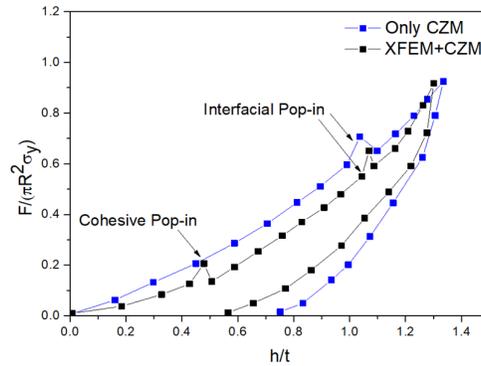

Figure 15. XFEM model of the nanoindentation process on c-Al0.66Ti0.33N / Interface / M2 system.

Fig. 16 shows the normalized crack length curve using the system thickness CIS vs. normalized force exerted by the indenter for each model. In this case, in both models a similar behavior regarding the cracking is observed. In this way, it is possible to see three cracking stages, these being crack nucleation, crack growth, crack opening and finally the stabilization. It is interesting to note that the cohesive crack, which is modeled with the CZM + XFEM model, nucleates before the interfacial crack, which is modeled with the CZM model. After the stage of opening crack, both types of crack are coupled and grow until their stabilization. This shows clearly the crack growth mechanisms in the CIS-system.

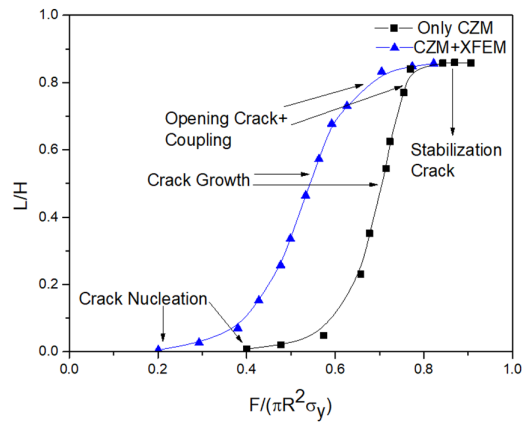

Figure 16. L / H vs. F / ($\pi R^2 \sigma_y$). L is the crack length, H is the thickness of the CIS system, F is the applied force, R is the radius of the indenter and $\sigma_y$ is the yield stress of the substrate. The stages of cracking and coupling between the two types of crack are observed.

## 4. Conclusions

A theoretical-experimental methodology for failure analysis of hard coatings is proposed. Properties associated to the fracture toughness of the coating material are experimentally obtained by means of indentation and scratch testing, and coupled with numerical XFEM models for understanding of cracking patterns induced by means of contact mechanics techniques.

The load vs. displacement data of the nanoindentation test showed that instantaneous fractures (*pop-in*) in the c-$Al_{0.66}Ti_{0.33}N$ arc-PVD coating form at 150 mN. Neither radial cracks (cohesive faults) nor medium cracks (adhesive faults) in the coating / interface / substrate system were found after applying loads between 0 and 300 mN. These results points out to the toughness of the coating deposited on the M2 steel. Prior cracking the nitride material plastically deforms (absence of radial cracks associated with cohesive faults). By means of the cohesive zone model and the extended finite element method, it was possible to determine the nucleation sites of the radial and lateral cracks, as well as the effect of their coupling in the system's cracking pattern. It was also possible to relate the load-displacement curve with the fractures through the analysis of the *pop-ins*. In this way, it was possible to validate the analysis methodology, showing that it is experimentally possible to obtain the values of the mechanical properties associated to the fracture toughness, which are necessary for the construction and implementation of numerical models associated with the finite elements method.

**Acknowledgements:**

The authors thank CONACyT for financial support provided through the program "Fronteras de la Ciencia" and the project 2015-02-1077. We also thank CONMAD, CENAPROT, LISMA and the scholarship given by CONACyT.